\documentclass[english,aps,prl,twocolumn, superscriptaddress,showpacs, amsmath]{revtex4-2}

\usepackage[utf8]{inputenc}
\usepackage{amsmath}
\usepackage{amsfonts}
\usepackage{graphicx}
\usepackage{soul}

\usepackage{natbib}
\usepackage{xcolor}
\usepackage{braket}
\usepackage{amssymb}
\usepackage{mathtools}
\usepackage{textcomp, gensymb}

\usepackage{graphicx,epsf,epsfig,ulem}
\usepackage{epsfig}
\usepackage{epstopdf} 
\usepackage{bm}
\usepackage{subfigure}
\usepackage{color}
\usepackage{physics}
\usepackage{tabularx}
\usepackage[T1]{fontenc}

\newcommand{\braketmatrix}[3]{\left \langle #1 \middle| #2 \middle| #3 \right \rangle}

\begin{document}

\preprint{APS/123-QED}

\title{Measurement of enhanced spin-orbit coupling strength for donor-bound electron spins in silicon}

\author{Radha Krishnan}
\author{Beng Yee Gan}
\affiliation{School of Physical and Mathematical Sciences,\\  Nanyang Technological University, 21 Nanyang Link, Singapore 637371}

\author{Yu-Ling Hsueh}
\affiliation{School of Physics, University of New South Wales, Sydney, NSW 2052, Australia}

\author{A.M. Saffat-Ee Huq}
\affiliation{School of Physics, University of New South Wales, Sydney, NSW 2052, Australia}

\author{Jonathan Kenny}
\affiliation{School of Physical and Mathematical Sciences,\\  Nanyang Technological University, 21 Nanyang Link, Singapore 637371}

\author{Rajib Rahman}
\affiliation{School of Physics, University of New South Wales, Sydney, NSW 2052, Australia}

\author{Teck Seng Koh}
\affiliation{School of Physical and Mathematical Sciences,\\  Nanyang Technological University, 21 Nanyang Link, Singapore 637371}

\author {Michelle Y. Simmons}
\affiliation{Centre for Quantum Computation and Communication Technology, School of Physics, University of New South Wales, Sydney, NSW 2052, Australia}

\author{Bent Weber}
\email{b.weber@ntu.edu.sg}
\affiliation{School of Physical and Mathematical Sciences,\\  Nanyang Technological University, 21 Nanyang Link, Singapore 637371}

\maketitle
\section{Abstract}
While traditionally considered a deleterious effect in quantum dot spin qubits, the spin-orbit interaction is recently being revisited as it allows for rapid coherent control by on-chip AC electric fields. For electrons in bulk silicon, SOC is intrinsically weak, however, it can be enhanced at surfaces and interfaces, or through atomic placement. Here we show that the strength of the spin-orbit coupling can be locally enhanced by more than two orders of magnitude in the manybody wave functions of multi-donor quantum dots  compared to a single donor, reaching strengths so far only reported for holes or two-donor system with certain symmetry. Our findings may provide a pathway towards all-electrical control of donor-bound spins in silicon using electric dipole spin resonance (EDSR).  

\section{Introduction}

Donor bound electron spins \cite{Watson2018, He2019atwo} in silicon are promising qubit candidates for the implementation of quantum information processing due to their exceptionally long spin relaxation ($T_1$) \cite{watson2017atomically} and coherence times ($T_2$) \cite{witzel2010electron}. This is owing to a low concentration of nuclear spins in the crystalline bulk matrix  \cite{witzel2010electron,veldhorst2014addressable}, especially in isotopically enriched $^{28}$Si, and generally weak coupling to charge noise due to the weak coupling of the electron's spin to orbital degrees of freedom \cite{zwanenburg2013silicon}. For donor spin qubits that are embedded deep within the silicon bulk crystal, far from any interface, inversion asymmetry of the silicon lattice precludes Dresselhaus spin-orbit coupling (SOC), and an isotropic Coulombic confinement potential gives rise to approximately spheroid-symmetric envelope wave functions in the lowest valley-orbit states, allowing for only weak Rashba SOC and higher-order electro-magnetic SOC \cite{weber2018spin}.

Traditionally, SOC has been considered as an undesired effect in spin qubits, as it can dominate spin relaxation \cite{song2017spin,scarlino2014spin,climente2013spin}, lead to state leakage through mixing of singlet ($S$) and triplet ($T^-$) states \cite{stepanenko2012singlet}, or an overall increased sensitivity to charge noise  \cite{ferdous2018interface}. In turn, however, SOC may be harnessed for rapid electrical control of individual spin qubits \cite{nadj2010spin}. Much effort has therefore recently been directed towards driving electric-dipole spin resonance (EDSR) transitions \cite{shafiei2013resolving,rashba2008theory} via the Stark shift \cite{ferdous2018valley} or by utilizing the disparity in the $g$-factors \cite{ruskov2018electron} for individual addressability of qubits via a microwave field  \cite{veldhorst2015spin,corna2018electrically,huang2017electrically,huang2019fidelity,krauth2022flopping}. Demonstrations have so far largely been restricted to hole spin qubits bound to silicon quantum dots \cite{wang2016anisotropic,li2015pauli} and acceptor atoms \cite{maurand2016cmos,van2018readout} in which SOC-mediated enhanced coupling to external AC electric fields \cite{bulaev2007electric,salfi2016quantum} may be combined with long spin life- \cite{bulaev2005spin} and coherence times \cite{testelin2009hole}. Enhanced SOC, however, has more recently also been reported for electron spin qubits in metal-oxide-semiconductor (MOS) quantum dots \cite{ferdous2018valley,corna2018electrically,harvey2019spin,tanttu2019controlling,li2015pauli}, as well as in donors \cite{weber2018spin, hsueh2024engineering}. However, the strength of the SOC ($\sim100$~neV) reported  \cite{ferdous2018valley,corna2018electrically,harvey2019spin,tanttu2019controlling,li2015pauli} is usually 2-3 orders of magnitude weaker for electrons compared to holes \cite{li2015pauli, van2018readout} (see Table~\ref{tab1}).

Several factors have been reported to enhance the weak SOC for electrons in silicon \cite{nestoklon2008electric,ferdous2018interface,weber2018spin}. For instance, electric fields at surfaces or interfaces, such as Si/SiO$_2$ \cite {jock2018silicon} or SiGe/Si \cite{nestoklon2008electric}, can enhance Rashba SOC \cite{lee2021synthetic}, while breaking of mirror symmetry and atomic-scale roughness of interfaces has been shown to produce a significant Dresselhaus SOC contribution \cite{ferdous2018interface}. In donor-based systems, strong electric fields can displace the electron wave function from the donor nucleus and can give rise to electric field-induced SOC \cite{weber2018spin}. Also, engineering two donor atoms at specific locations can change local symmetry and result in both Rashba and Dresselhaus SOC \cite{hsueh2024engineering}.For hole spins, SOC is known to be strongly enhanced, owing to the reduced symmetry of the non-spheroid spin carrier wave function \cite{liles2018spin}.

\begin{figure*}[t]
\begin{center}
  \includegraphics[scale=0.5]{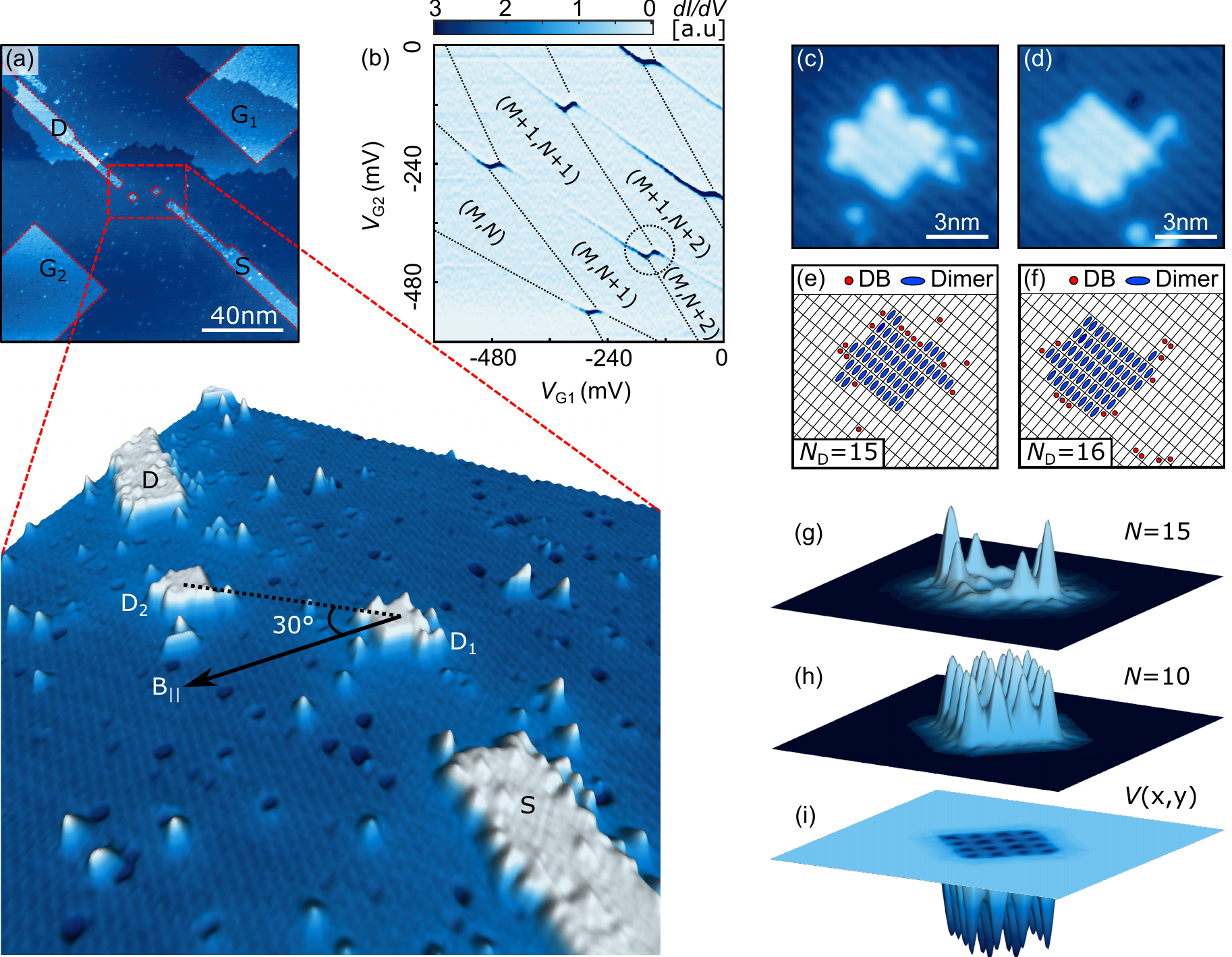}
 \caption[\textbf{Si:P DQD Device.}]{\textbf{Si:P DQD Device.} \textbf{(a)} Overview STM-image after hydrogen resist lithography, showing two quantum dots ($D_{1,2}$), connected to source (S), drain (D), and gate ($G_{1,2}$) electrodes. \textbf{(b)} The charge stability diagram of the DQD shows a number of closed honeycomb domains, where $(M,N)$ indicate the effective electron occupation. \textbf{(c,d)} Atomic-resolution STM images of the two quantum dots allowing to estimate the approximate number of incorporated P donors \textbf{(e,f)}. \textbf{(g-i)} Self-consistent confinement potential (i) and tight-binding wave-functions, calculated for the $N=15$ (g) and $N=10$ (h) charge state in NEMO-3D.}
 \label{fig1}
 \end{center}
\end{figure*}

Similarly, as we show in this work, SOC can be significantly enhanced by more than 2 orders of magnitude for electrons in multi-donor quantum dots due to an interplay of the atomically-sharp confinement potential with dopant placement disorder, a small valley-orbit splitting and non-spherical manybody wave functions. We draw our conclusions from the transport spectroscopy of a multi-donor double quantum dot in the Pauli spin blockade regime, in which a pronounced leakage is observed at the degeneracy point of the effective valence spin singlet $S(0,2)$ and triplet $T^-(1,1)$ states. A transport model of spin-dependent second-order tunneling confirms that this resonance can only be explained by an SOC mediated coupling of the singlet and triplet states, while atomistic tight-binding calculations \cite{klimeck2007atomistic} confirm the magnitude of the SOC observed.    

\begingroup
\setlength{\tabcolsep}{6pt} 
\begin{table}[ht]
\centering
\caption{\textbf{Spin-orbit interaction strength for electrons and holes in silicon.}}
\label{tab1}
\begin{tabular}{l c c}
\textbf{Reference} & \textbf{Carrier}    & \textbf{SOC strength}\\ \hline
Harvey-Collard et al.~\cite{harvey2019spin}    & Electron   & (113$\pm$22)~neV  \\
Tanttu et al. \cite{tanttu2019controlling}    & Electron   & $\lesssim$ 113.73 neV   \\
Jock et al. \cite{jock2018silicon}    & Electron   & $\lesssim$ 82.7 neV  \\
Weber et al. \cite{weber2018spin}    & Electron   &$\lesssim$ 1.05 $\mu$eV      \\
Hsueh et al. \cite{hsueh2024engineering}   & Electron& 33.9$\mu$eV\\
Li et al. \cite{li2015pauli}   & Hole      & 110 $\mu$eV   \\
Heijden et al. \cite{van2018readout}   & Hole      & (36$\pm$5) $\mu$eV\\

\bf{This Work} & Electron & (40$\pm$10) $\mu$eV \\
\end{tabular}
\end{table}
\endgroup

\section{Results and Discussion}

The multi-donor double quantum dot device is shown in Fig.\ref{fig1}a, fabricated by STM hydrogen resist lithography \cite{weber2012engineering}. Both quantum dots $D_1$ and $D_2$ are $\sim$4~nm in diameter, separated by $17.5 \pm 0.5$~nm  (centre to centre), each containing approximately $N_D \simeq 15$ phosphorus (P) donors \cite{weber2012engineering} as estimated from the size of the lithographically defined dots (Fig.~\ref{fig1}c-f), assuming that six contiguous dimers of the Si(001)-2$\times$1 surface are required to incorporate a phosphorus atom into the silicon matrix \cite{fuechsle2012single}. Both dots are mutually tunnel-coupled and are coupled to source (S) and drain (D) electrodes which have been staggered to allow independent electrostatic control of the individual dots' electrochemical potentials by gate electrodes G$_1$ and G$_2$ \cite{weber2012engineering}. A DC magnetic field was applied within the plane of the device, along the Si$\left<110\right>$ direction, i.e. $\sim 30 \degree$ from the axis connecting the two quantum dots (arrow in Fig.~\ref{fig1}a).

Figure~\ref{fig1}b shows the measured charge stability diagram, indicating electron numbers $N$ and $M$ per dot. Absolute electron numbers cannot be determined as the dots could not be fully depleted within the accessible gate-voltage range. However, we expect $N,M \approx N_{\rm D} \simeq 15$ near charge neutrality ($V_{\rm G1} = V_{\rm G2} = 0$~V) \cite{weber2012engineering, weber2014spin}. 

\begin{figure*}[t]
\begin{center}
\includegraphics[scale=0.6]{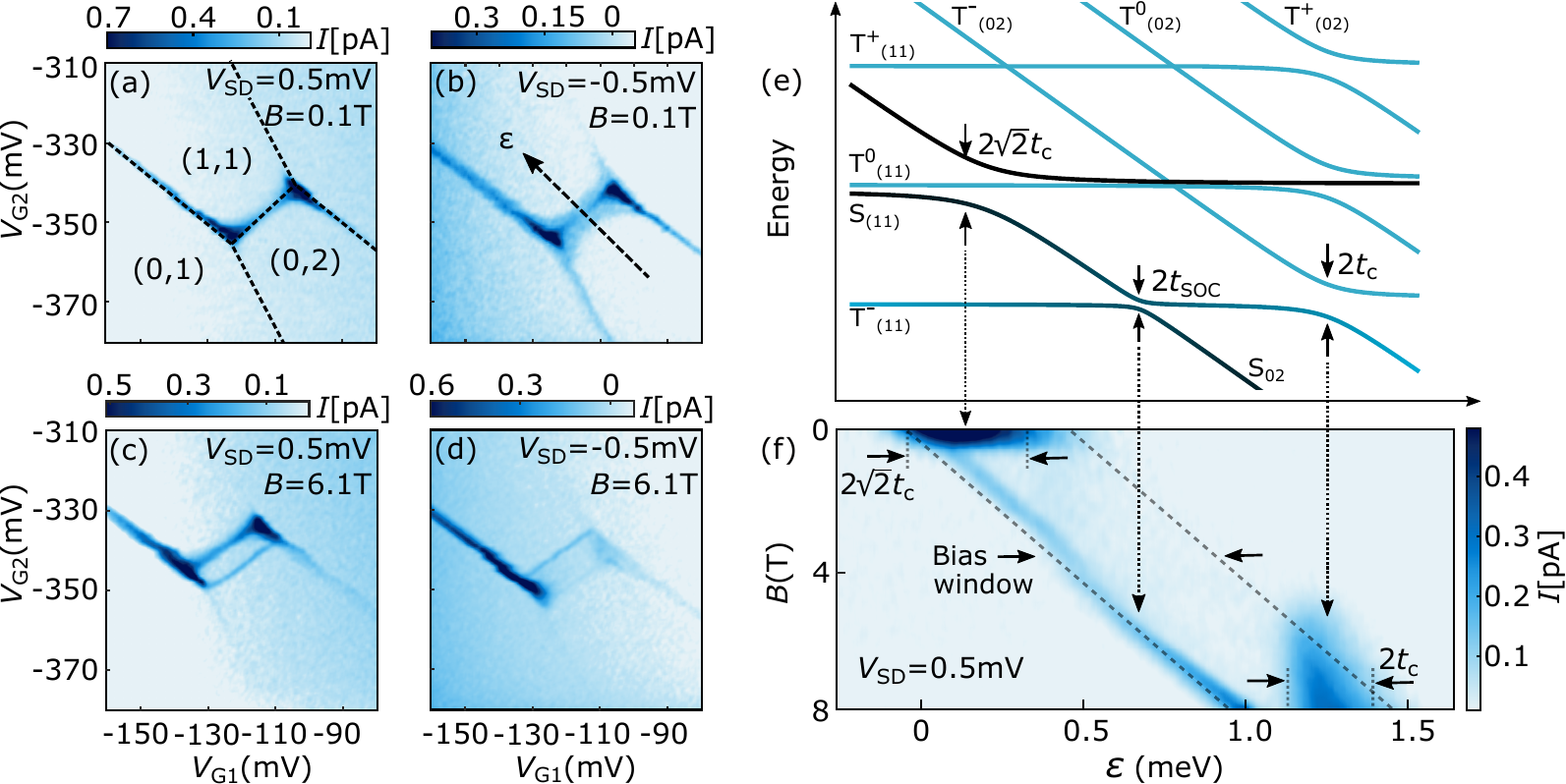}
 \caption[\textbf{Low-bias spectroscopy of the Si:P DQD Device.}]{\textbf{Low-bias magnetospectroscopy of the Si:P DQD Device. (a-d)} Close-up of the charge transition highlighted by the dashed circle in Figs.~\ref{fig1}b, measured at $V_D= \pm 500 \mu V$, and at $B = 0$~T (a,b) and $B=6.1$~T (c,d), respectively. \textbf{(e,f)} Co-tunneling current, measured along the dashed arrow in (b) and plotted as a function of magnetic field and detuning, compared to the corresponding energy eigenspectrum (e) of effective two-electron singlet and triplet states. $t_{\rm c}$ denotes the tunnel coupling, and $t_{\rm SOC}$ the spin-orbit coupling. The bias window is indicated in (f) and the data inverted in $B$ for ease of comparison with (e).}
\label{fig3}
\end{center}
\end{figure*}
	
Despite the large electron numbers, we observe well-defined effective two-electron singlet and triplet states as summarized in Fig.~\ref{fig3}. A close-up of the charge transition highlighted by the black dashed circle in Fig.~\ref{fig1}b is shown in Figure~\ref{fig3}a-d, recorded at a low bias of $V_{SD} = \pm 500~\mu$V and at two different values of the applied in-plane magnetic field ($B=0$~T and $B=6.1$~T). At larger bias, the charge degeneracy (triple) points widen into bias triangles \cite{weber2012engineering}, exhibiting conventional Pauli spin blockade (see Supporting Information). Current along the outlines of the charge stable domains at $B=100$~mT (dashed white lines) arises from higher-order (co-)tunneling. 

Current resonances connecting the triple points reflect co-tunneling due to a strong mutual tunnel coupling between the two quantum dots. At finite magnetic fields $B = 6.1$~T (Fig.~\ref{fig3}c,d), we observe that the single inter-dot co-tunneling resonance splits into two. This is more clearly observed in the evolution of the co-tunneling current resonances as a function of magnetic field, recorded along the dashed arrow in Fig.~\ref{fig3}b, and plotted in Fig.~\ref{fig3}f. The corresponding energy eigenspectrum of SOC-perturbed effective $(1,1)$ and $(0,2)$ singlet- ($S$) and triplet- ($T$) valence spin states \cite{tarucha1996shell} is shown in Fig.~\ref{fig3}e. For ease of comparison, the data was rescaled with respect to detuning energy $\varepsilon$, assuming a lever arm ($\alpha = 0.08 \pm 0.01$), extracted from the observed bias window ($500~\mu$eV), and with the magnetic field axis inverted. We define zero detuning ($\varepsilon =0$) where the singlet states $S(1,1)$ and $S(0,2)$ are degenerate, and hybridize due to the tunnel coupling $t_c$. This is visible from a high intensity resonance at low magnetic fields ($B < 1$~T), that is symmetric upon bias reversal. 

\begin{figure*}[t]
\begin{center}
  \includegraphics[width=0.9\textwidth]{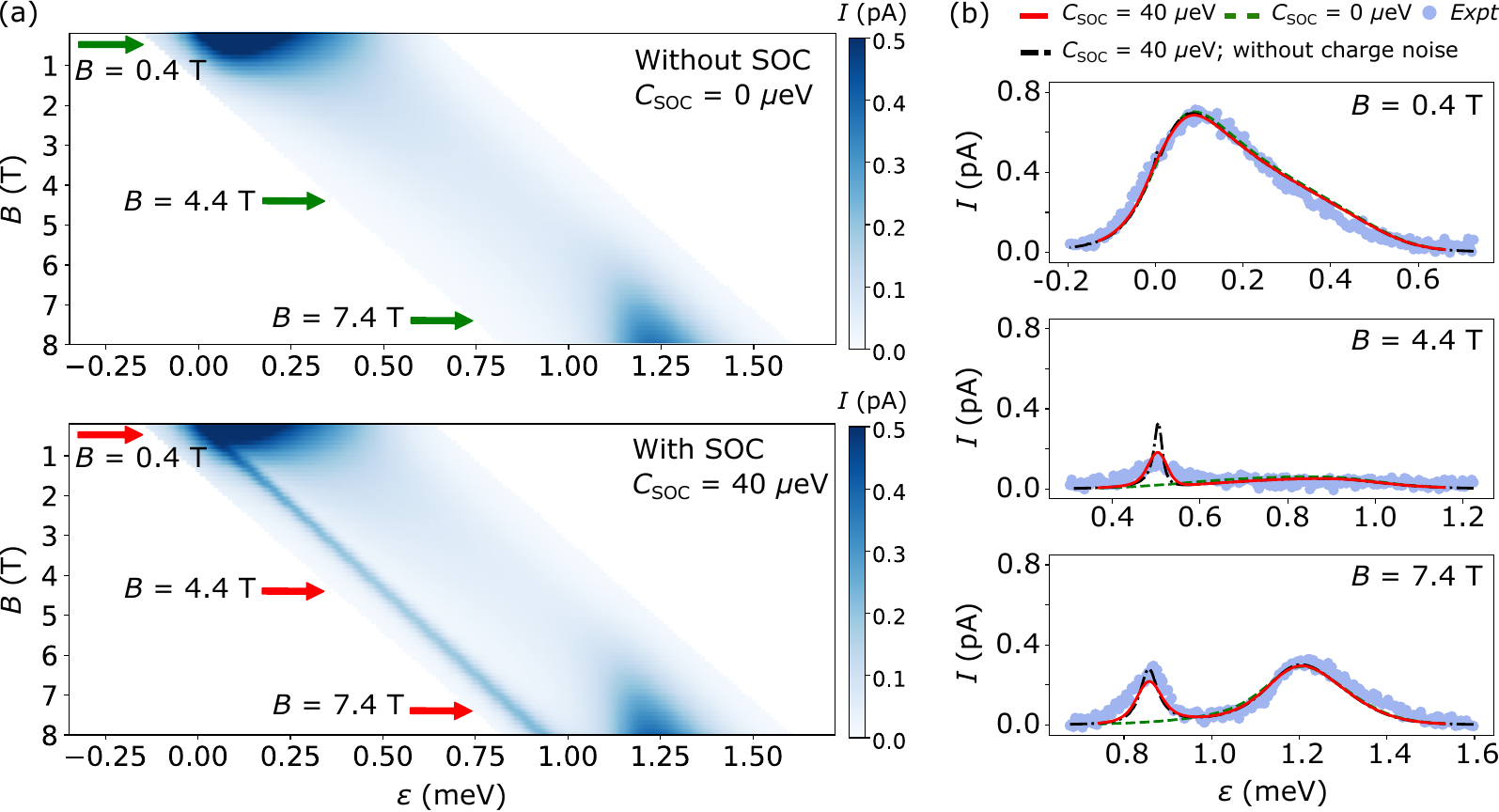}
  \caption[\textbf{Effects of SOC in second order transport.}]{\textbf{Effects of SOC in second order transport.} (a) Second order transport model of the data in Fig.~\ref{fig3}, with and without SOC-mediated coherent coupling included (see main text and Supporting Information for detail). No significant co-tunneling current is observed when SOC-mediated coupling is absent while a clear $T^-(1,1)-S(0,2)$ transport resonance is observed in the presence of SOC. (b) Line cuts comparing experiment (Fig.~\ref{fig3}f) and theory in terms of resonance line widths and intensities of for low, intermediate and high fields, as indicated by the red arrows in (a). Red and green lines show our results at $C_\text{SOC} = 40~\mu\mathrm{eV}$ and $0~\mu\mathrm{eV}$, respectively, and with Gaussian (charge) noise of magnitude $\sigma_\varepsilon = 17~\mu$eV \cite{Kranz2020exploiting}) in the detuning parameter. We see that the charge noise has the effect of broadening and reducing the intensity of the $S(0,2)-T^-(1,1)$ resonance, as can be seen by comparing to results without charge noise (black dashed line).}
\label{fig4}
\end{center}
\end{figure*}

At finite magnetic field, the degeneracy of the $T(1,1)$ and $T(0,2)$ triplet manifold is lifted, resulting in three non-degenerate levels each, split by the Zeeman energy, $\Delta E_Z = S_z g \mu_B B$. Positive detuning lowers the energy of the $(0,2)$ states, allowing the triplet states to hybridize at $\varepsilon = 1.21$~meV. This resonance becomes only visible at $B\sim 5$~T, when the hybridized $T^-$ states enter the bias window (horizontal arrows in Fig.~\ref{fig3}f). As the detuning at which level alignment of $S(1,1)-S(0,2)$ and $T(1,1)-T(0,2)$ states occurs does not depend on magnetic field, the corresponding transport resonances do not change position. The finite magnetic field also leads to a change in the $(1,1)$ ground state from singlet to triplet, giving rise to a level crossing of $T^-(1,1)$ and the $S(0,2)$ states when $\Delta E_Z > t_c$ at $B \simeq 1$~T. An enhanced occupation probability of the $T^-(1,1)$ over the $S(1,1)$ result in a suppression of co-tunneling current into the drain lead, and can be understood as a manifestation of Pauli spin blockade in the co-tunneling regime \cite{liu2005pauli}. Current at $B \lesssim 1$~T, also reflected in our transport calculations (Fig.~\ref{fig4}), likely has its origin in spin-flip co-tunneling with the S lead \cite{coish2011leakage, lai2011pauli}.

At fields $B \gtrsim 1$~T, spin selection rules should prohibit co-tunneling at the $T^-(1,1) - S(0,2)$ level crossing. Yet, a fainter but sharp and clear current resonance is observed that shifts linearly as a function of magnetic field. Such current resonance can only be explained by the presence of a spin non-conserving coupling, such as inelastic co-tunneling \cite{liu2005pauli, moriyama2009inelastic}, nuclear hyperfine interaction \cite{borhani2010two, johnson2005triplet, petta2005coherent, laird2006effect, petta2008dynamic}, or, spin orbit coupling  \cite{song2017spin, harvey2019spin, meunier2007experimental, amasha2006measurements, scarlino2014spin}.

Such coupling of $T^-(1,1) - S(0,2)$ states as observed can be understood from the formation of spin-orbit perturbed valley-orbit eigenstates \cite{yang2013spin, corna2018electrically}. From first-order perturbation theory, the valley-states in each QD can be expressed as admixtures of both spin polarities by virtue of the spin-orbit coupling,

\begin{eqnarray}\label{spin-states1}
    \ket{\Downarrow} = \ket{v_i, \downarrow} - \frac{C_{\rm SOC}}{\Delta E + \tfrac{1}{2}g\mu_B B} \ket{v_j, \uparrow} + ...
\end{eqnarray}
and 
\begin{eqnarray}\label{spin-states2}
    \ket{\Uparrow} = \ket{v_i, \uparrow} - \frac{C^*_{\rm SOC}}{\Delta E - \tfrac{1}{2}g\mu_B B} \ket{v_j, \downarrow} + ...
\end{eqnarray}

Here, $E_Z = g\mu_B B$ is the Zeeman energy with the electron $g$-factor $g \approx 2$, and $\Delta E$ is the valley-orbit splitting. $C_{\rm SOC} = \braketmatrix{v_j,\uparrow}{H_{\rm SOC}}{v_i,\downarrow}$ is the SOC matrix element. The degree of admixture of the spin-valley states thus sensitively depends on the magnitude of the valley-obit splitting $\Delta E$. Naturally assumed to be large in atomically confined systems, the valley multiplicity in silicon, especially at large charge occupation \cite{fuechsle2010spectroscopy, weber2012engineering}, provides for comparatively small valley-orbit gaps in the range of a few hundred $\mu$eV \cite{fuechsle2010spectroscopy} to a few meV \cite{weber2012engineering}, and thus comparable in magnitude to the much larger few-electron quantum dots realized in SiGe or Si MOS \cite{yang2013spin,corna2018electrically}. 

Given that the spin-valley states of the individual quantum dots contain admixtures of both the spin polarities, they further give rise to a coherent coupling between singlet and triplet states across the two quantum dots of magnitude $C_\text{SOC}/\sqrt{2}$. At intermediate magnetic fields, where these are ground states, the magnitude of the effective coupling can be estimated from a Schrieffer-Wolff transformation, as

\begin{equation}
    t_\text{SOC}= t_c C_\text{SOC}/\sqrt{2} \left| \tfrac{1}{\varepsilon - g\mu_B B}-\tfrac{1}{\varepsilon - \Delta E} \right |
\end{equation} 

This is hence a second order coupling in which $S(0,2)$ is coupled to $T^-(0,2)$ via SOC in the occupied dot, and $T^-(0,2)$ to $T^-(1,1)$ via tunneling.

We confirm this notion from a transport model in the co-tunneling regime as summarized in Fig.~\ref{fig4} (see Supporting Information for details). Our model reproduces all essential features of the data, but most notably the presence of a single pronounced co-tunneling resonance at intermediate magnetic fields, observed only when the SOC-mediated coupling between the dots is present. This is clearly reflected in the colorplots of Fig.~\ref{fig4}a as well as the individual line traces (Fig.~\ref{fig4}b) at three different strengths of the magnetic field, comparing the measured drain current with that calculated in second-order transport.

We note that the overall current intensity is dominated by the product of dot-lead tunnel rates $\Gamma_S$ and $\Gamma_D$, while the relative intensities of the respective resonances depend non-linearly on these rates. For example, the current intensity at the singlet resonance observed at low magnetic field ($< 1$~T) is strongly dependent on dot-lead tunnel rates as it is determined through steady-state DQD populations. Spin blockade occurs when the DQD is in a triplet (1,1) state, lifted by co-tunneling of a spin between a dot and the adjacent lead. Such spin non-preserving processes, in particular spin-flip co-tunneling \cite{coish2011leakage}, are therefore implicitly taken into account in the model, but can only explain strong resonant current at $B <1$~T. For $B > 1$~T, the onset of Pauli spin blockade sets an upper bound on $t_c$.

Our model also confirms that the resonance lineshapes are Lorentzian when lifetime broadening is dominant. In this case, linewidths and peak intensities are determined by the magnitude of $t_c$ and $C_\text{SOC}$. Interestingly, however, the $S(0,2)-T^-(1,1)$ resonance also possesses a Gaussian character which suggests an inhomogeneous broadening mechanism, for instance due to charge noise in detuning. We can model this by convolving the calculated current (see Supporting Information) with a Gaussian of half-width $\sigma_\varepsilon = 17~\mu$eV, in reasonable agreement with recent measurements of bandwidth-integrated charge noise in epitaxial Si:P devices \cite{Kranz2020exploiting, He2019atwo}. We note that the current lineshape through the $T^-(1,1)-S(0,2)$ channel is under-determined if both $t_c$ and $C_\text{SOC}$ are unbounded. However, the upper bound on $t_c$ constrains $C_\text{SOC}$ from below, which can then be estimated. In a similar way, the upper bound on $t_c$ allows the energy splitting $\Delta E$ to be estimated from the lineshape and position of the triplet channel.

\begin{figure}[t]
\begin{center}
  \includegraphics[scale=0.5]{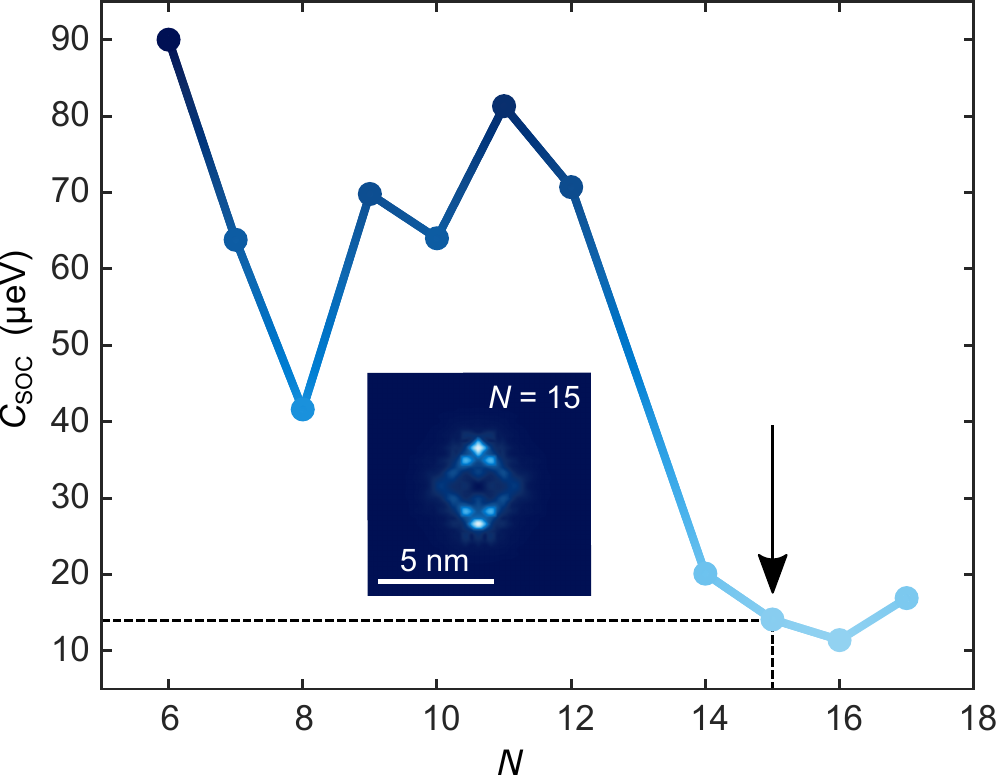}
  \caption[\textbf{Calculated SOC strength.}]{\textbf{Calculated SOC strength.} Strength of the spin-orbit coupling $C_\textrm{SOC}$ as calculated from atomistic tight-binding, and plotted as function of the electron number $N$. The decrease of the SOC strength towards higher electron numbers likely reflects screening of the donor-based Coulombic confinement potential. Insert shows the envelope wavefunction for N =15.}
\label{fig2}
\end{center}
\end{figure}

From fits to the data, we extract $C_\text{SOC} \approx 40~\mu\mathrm{eV}$, more than two orders of magnitude larger than values previously reported for electrons in silicon, reaching values previously only observed for holes (compare Table~\ref{tab1}). Such considerable enhancement of the SOC strength likely arises as a combination of several factors, including the atomically-abrupt and disordered Coulombic  confinement potential of the dots (Fig.~\ref{fig1}h) lowering wave function symmetry and hence increasing the symmetry dependent SOC transition dipole moment \cite{tahan2014relaxation}. Indeed, we can confirm the magnitude of the SOC strength by atomistic tight-binding calculations of the multi-donor dots, as shown in Fig.~\ref{fig2}, plotting $C_\textrm{soc}$ against electron occupation $N$. Sixteen P donors, represented by Coulomb charges each with a cutoff potential capturing the central-cell correction, are placed in the Si crystal within a 4 by 4 nm region. The tight-binding Hamiltonian is set up using 20 atomic orbital sp3d5s* basis including nearest-neighbor interactions \cite{klimeck2007}. A self-consistent Hartree method is used to capture the multi-electron wavefunction, which has successfully reproduced the two-electron binding energy in a single donor \cite{rahman2011electric}. The SOC estimates have been extracted from the self-consistent TB wave functions, directly, by considering calculations with spin-orbit ($\ket{\Uparrow}$, $\ket{\Downarrow}$) and without ($\ket{v_i,\uparrow}$,  $\ket{v_j,\downarrow}$) by setting the silicon tight-binding spin-orbit parameter to the bulk value \cite{chadi1977spin, boykin2004valence} or zero, respectively. A projection from the spin-orbit wave function onto the pure orbital basis then gives us $C_{\rm SOC}$ as plotted in Fig.~\ref{fig2}, reaching values between 11~$\mu$eV ($N=16$) and 90~$\mu$eV ($N=6$), matching the value extracted from the experiments (40~$\mu$eV) in their order of magnitude. For a range of different disordered donor configurations within the dots, we find values $\sim$10-20~$\mu$eV at $N \simeq 15$, confirming that these results are relative robust against dopant placement disorder within the dots. While the model slightly underestimates the SOC strength at $N \simeq 15$, a strong enhancement towards lower electron occupation likely reflects the less effective screening of the Coulombic confinement potential, amplifying the effects of donor disorder and wave function asymmetry.   

\section{Conclusion}
To conclude, we have reported a strong enhancement of the spin-orbit coupling (SOC) strength for spins in the manybody wavefunctions of multi-donor silicon quantum dots, with an SOC energy scale as much as $40~\mu$eV -- exceeding previous reports for electrons in silicon quantum dots by two orders of magnitude. We explain this enhancement by the abrupt and disordered nature of the Coulombic confinement potential inherent to donor-based quantum dots, reducing wave-function symmetry. Such strongly enhanced SOC in silicon -- similar in magnitude to that for holes -- may provide a pathway towards all-electrical control of donor-bound spins in silicon by electric dipole spin resonance (EDSR) as well as may enhance coupling of donor spins to superconducting resonators.        

\section*{Acknowledgements} 
This research is supported under the Singapore Quantum Engineering Programme (QEP2.0) ``Atomic Engineering of Donor-based Spin Qubits in Silicon'' (NRF2021-QEP2-02-P07). BW acknowledges a Singapore National Research Foundation (NRF) Fellowship (NRF-NRFF2017-11).

\bibliography{References}
\bibliographystyle{ieeetr}

\end{document}